\documentclass[%
preprint,
 amsmath,amssymb,
 aps,
pra,
]{revtex4-1}

\usepackage{graphicx}
\usepackage{dcolumn}
\usepackage{bm}
\usepackage{nicefrac}

\begin{document}


\title{Quantum Analytical Mechanics: Quantum Mechanics with Hidden Variables}

\author{Wolfgang Paul}%
 \email{Wolfgang.Paul@physik.uni-halle.de}
\affiliation{%
Institut für Physik, Martin-Luther Universität Halle-Wittenberg\\
06099 Halle, Germany
}%

\date{\today}

\begin{abstract}
The question about the existence of so-called ``hidden'' variables in quantum
mechanics and the perception of the completeness of quantum mechanics are two
sides of the same coin. Quantum analytical mechanics constitutes a completion
of standard quantum mechanics based on the concept of stochastic trajectories
in the configuration space of a quantum system. For particle systems,
configuration space is made up out of their coordinates and, if relevant,
their orientation. Quantum analytical mechanics derives equations of motion
for these variables which allow a description of the measurement process as a
dynamical physical process. After all, it is exactly these variables
experiments are designed to interact with. The theory is not a replacement of
Hilbert space quantum mechanics but a mathematical completion enriching our
toolset for the description of quantum phenomena.
\end{abstract}

\maketitle

\section{Introduction}
During the 5th Solvay Conference in 1927 Bohr and Einstein reportedly had many
breakfast and dinner discussions about the question whether the mathematical
formalism of quantum mechanics, as it was developing at that time, could be
considered complete \cite{Guido-2009}. Einstein did not accept that and called what was 
missing ``hidden'' variables. ``Missing'' variables would have been a better
choice of words (we will, however, continue to use the term ``hidden'' in the
following), as it turns out that these variables are actually the only 
thing not hidden in experiments. No experiment has ever or will ever be done
in Hilbert space, but standard Hilbert space quantum mechanics contains no
elements able to describe the measurement process dynamically. Such a description of the
measurement process was declared unattainable in Copenhagen quantum mechanics
as encoded in the Dirac-von Neumann axioms. At first it seemed that Bohr had won the debate with Einstein
\cite{Heisenberg29}, especially when John von Neumann published his book
\emph{Mathematische Grundlagen der Quantenmechanik} \cite{JvN}. Besides
defining the mathematical structure of Hilbert space quantum 
mechanics, this book contained what has been called the ``no hidden variables theorem''. The
theorem aimed to show that when one assumes that one can add further variables
to the quantum mechanical description leading to a \emph{dispersion free
  ensemble}, this assumption is incompatible with the mathematical structure
of the theory.

The perception on who won the 1927
dispute should have been short-lived, as already in 1935 Grete Herrmann
published a work \cite{Grete35} showing that von Neumann's proof was based on an
assumption not realisable in physics. He needed to assume that the sum of two
non-commuting Hermitian operators, which each correspond to a physical
observable, is not only a new Hermitian operator but also a physical observable.
Unfortunately, this work was completely ignored in the physics community, and,
consequently, (almost) all our textbooks, which began to appear since then, are
based on the Dirac-von-Neumann axioms and the Copenhagen interpretation of
quantum mechanics. With these 90 years of history, most physicists today
will take it as an established scientific fact, that there is no way to obtain
a description of the measurement process in terms of a physical dynamical
theory. The fact that this belief did cut off one of the legs on 
which physics had stood for 300 years, and left book-keeping of the outcome
statistics as the only goal of quantum physics, seems not to worry most physicists
any longer. This is different among philosophers of physics because, in a
sense, the Copenhagen interpretation declared the ``end of the scientific
project to deliver a communicable picture of physical reality''
\cite{Albert}. I will discuss that, as Einstein expected, this reduction of
the scope of physics was premature. I will present a perspective on the
question of hidden variables in quantum mechanics and how quantum analytical
mechanics may solve this long standing problem. 

\section{A Short(ened) History of Hidden Variables}
In 1952, David Bohm \cite{Bohm-1952,Bohm-1952a} suggested a hidden variable
theory rediscovering an earlier suggestion by Louis de Broglie
\cite{Louis}. This theory is completely in agreement with Hilbert space
quantum mechanics as far as the prediction of measurement statistics is
concerned. It has been developed into a self-contained (statistical)
mechanical theory by now \cite{Duerr}. Thus, at the end of the 1950s, there
existed a hidden variable theory of quantum mechanics and the general conviction, that von
Neumann had proven that no such thing was possible. This motivated the work of
John Stewart Bell who published a manuscript showing the flaw in von Neumann's proof
\cite{Bell-1966}. He recognised the same problem that Grete Herrmann had discussed
30 years before.

But Bell went on and derived an inequality which a realistic,
local hidden variable theory needed to fulfil \cite{Bell-1964}, based on some
formal assumptions that he felt unavoidable for such a theory. This 
inequality is violated by entangled states in Hilbert space quantum
mechanics. The fact that this prediction was experimentally testable led to 50
years of careful experiments proving that nature, in fact, violates Bell-type
inequalities, which was rightfully honoured with the Nobel prize in physics of
2022 \cite{nobel2022}. Unfortunately, the Nobel committee in its laudation \cite{nobel2022}
concluded that \emph{...if there are hidden variables, the
correlation between the results of a large number of measurements  will never
exceed a certain value.} This statement ignores the logical content
of Bell-like theorems and is therefore plain wrong. Hilbert space quantum
mechanics of entangled states violates the assumption of Bell separability
\cite{Hall-2016}, which assumes the factorization of the joint probability for
the outcome of, e.g., measurements of the spin projections of two particles in
an entangled state. This is at the origin of the violation of Bell-type
inequalities. This non-separability is often equated with non-locality, which
is logically not the same, and it is concluded that a hidden-variable theory
in accord with Hilbert space quantum mechanics has to be non-local.

\subsection{Stochastic Completions of Quantum Mechanics}
Also in the year 1927, Heisenberg suggested his famous thought experiment of
the gamma ray microscope to provide a physical understanding of the
uncertainty principle. Scattering of a single photon is enough to change the
state of motion of, e.g., an electron to a significant degree. However, he and
the other fathers of matrix  mechanics built their theory on the concept that,
when no external observer tries to scatter a photon off the electron, it can be considered
as an isolated mechanical system. This completely ignored what was known about the omnipresent
black body radiation, the actual birth place of quantum physics. Based on Heisenberg's
thought experiment it would have been more natural to seek a description of
quantum systems as open systems, however, there existed only dissipative
theories of open systems at that time, which were incompatible with the
well-established fact of energy conservation in the quantum world. So model building in quantum
mechanics was designed following the procedure of classical mechanics to
isolate the objects of interest (e.g., proton and electron in a hydrogen atom)
and take all interactions with the rest of the universe into account
perturbatively. Heisenberg's thought experiment said, that this was actually
a - to say the least - highly questionable procedure. 

Schrödinger perceived early on that his theory
\cite{Schroedinger-I,Schroedinger-II,Schroedinger-III,Schroedinger-IV,Schroedinger-V} has
stochastic aspects \cite{Schroedinger-1931} and instead of the Schrödinger equation
one could consider one diffusion equation forward in time and another one
backward in time. This gave rise to the mathematical theory of Bernstein
processes \cite{Bernstein} as well as the study of so-called Schrödinger
bridges, reversibly connecting probability distributions in Wasserstein space,
from which one can actually derive the Schrödinger equation
\cite{Renesse}. Schrödinger's euclidean version of a stochastic quantum theory
was strongly pursued in the work of Zambrini \cite{Zambrini-1987} and
collaborators. A variation of this approach lead to the quantum analytical
mechanics, which we will discuss in more detail in the next section. 

First let us state, that today there seems to be a widespread accord, that a
completion of quantum mechanics has to be done in the form of a stochastic
theory. In a series of recent papers, Jacob Barandes
\cite{Barandes-2025,Barandes-2024} has shown that Hilbert 
space quantum mechanics arises naturally for a stochastic system
following an indivisible stochastic process on its configuration
space. This indivisibility basically 
means that one does not need to assume more than the existence of two
stochastic events linked by a conditional probability matrix $\Gamma_{ij}(t)$ (so far, the
theory is defined as a stroboscopic dynamics on a discrete event space). For
unistochastic matrices, this conditional probability matrix always can be
written as the modulus squared of a unitary matrix $\Gamma_{ij}(t) = \left |
  U_{ij}(t) \right|^2$. The unitary matrix gives rise to the unitary time
evolution of a state vector, i.e., the Schrödinger equation. This
approach is very general and extremely minimalistic, yet it allows for a
prediction of almost all the phenomenology of quantum mechanics. Quantum
analytical mechanics, in contrast, is a Markov theory, but it is concerned
with deriving the unitary time evolution as a part of the mechanical theory,
so it is not in conflict with the indivisibility on the phenomenology side.
A stochastic mapping between discrete events also underlies the completion
of quantum mechanics Jürg Fröhlich and collaborators have suggested over the last ten years
\cite{froehlich-2016a,froehlich-2021,froehlich-2025,froehlich-2025a}. They
consider a stochastic process on the algebra of 
physical observables, where each event is defined through a dissipative interaction with
the environment, e.g., preparation of a system, a measurement performed on the
system, but also the spontaneous decay of a nucleus and similar events. These
events branch into trees and define the stochastic history of a quantum
system. While Fröhlich considers specifically a representation of physical
observables by Hermitian operators, the concepts could as well be applied to
stochastic variables on configuration space as the realisation of the algebra
of observables, linking this approach to the one by Barandes and the one we
will discuss in the next section.

\section{Quantum Analytical Mechanics}
In the year 1966, Edward Nelson \cite{Nelson-1966}, building on earlier ideas of
Fenyes \cite{Fenyes}, developed a Newtonian mechanics for time inversion
invariant diffusion processes as a model for the dynamics of quantum systems. 
The kinematics of this Newtonian dynamics is defined through two diffusion
equations, one forward in time and one backward, which define the same process
in probability (here written for d=1 for simplicity; $m$ is the mass of the particle ) 
\begin{eqnarray}
 dx(t) &=& \left[v(x(t),t) + u(x(t),t) \right] dt + \sqrt{\frac{\hbar}{m}} dW_f(t)\\
 dx(t) &=& \left[v(x(t),t) - u(x(t),t) \right] dt + \sqrt{\frac{\hbar}{m}} dW_b(t)\;.\nonumber
\end{eqnarray}
Here the first equation is forward in time, with a forward in time Wiener
increment $dW_f(t)$ independent of all positions $x(s), s\ge t$, and the
second one is backward in time with a Wiener increment $dW_b(t)$ independent
of all positions $x(s), s\le t$. The probability streaming velocity is the
derivative of the action $v(x,t)=\nicefrac{1}{m}\,\partial_x S(x,t)$ and the osmotic
velocity is the derivative of the logarithm of the probability density of the
process $u(x,t)=\nicefrac{\hbar}{2m}\, \partial_x\ln[\rho(x,t)]$. 
Energy conservation in the mean is ensured (for conservative forces) through a
stochastic Nelson-Newton law 
\begin{equation}
 \frac{m}{2}\left[D_f D_b + D_b D_f\right]x(t) = F(x(t))\;,
\end{equation}
with the derivations defined as expectation values
\begin{eqnarray}
D_f f(x(t),t) &=& \lim_{\Delta t \to 0+}  \mathbb{E}\left[\frac{f(x(t+\Delta t),t+\Delta t) - f(x(t),t)}{\Delta t}\right]\\
D_b f(x(t),t) &=& \lim_{\Delta t \to 0+} \mathbb{E}\left[\frac{f(x(t),t) - f(x(t-\Delta t),t-\Delta t)}{\Delta t}\right]\nonumber
\end{eqnarray}
This construction is necessary because the solutions $x(t)$ of the kinematic
equations are everywhere continuous but nowhere differentiable. Nelson was
able to  derive the Schrödinger equation, or, more 
precisely, the Madelung equations, as the Hamilton-Jacobi formulation of this
stochastic Newtonian mechanics. 
\begin{eqnarray}
 \partial_t \rho(x,t) + \partial_x\left[ \frac{1}{m}\partial_xS(x,t) \rho(x,t)\right] &=& 0\\
 \partial_t S(x,t) + \frac{1}{2m}\left(\partial_x S(x,t) \right)^2 +V(x) -\frac{\hbar^2}{2m}\frac{\partial_x^2\sqrt{\rho(x,t)}}{\sqrt{\rho(x,t)}} &=&0\nonumber
\end{eqnarray}
Nelson already hypothesized that the fluctuations might originate in a coupling to the
electromagnetic background radiation. This was theoretically developed in the
group of de la Pe{\~n}a-Auerbach and Cetto \cite{Pena-Cetto-book} as a coupling
to the zero-point fluctuations of the electromagnetic field.

Thirty years after Nelsons original work, Michele Pavon
\cite{Pavon-1995,Pavon-1995a} presented the definitive 
quantum version of the Hamilton principle of classical mechanics. He included
the two velocity fields governing the kinematics of the diffusion processes
into what he called the quantum velocity $v_q(x,t) = v(x,t) - i u(x,t)$.
\begin{equation}
J[v_q]= \mbox{extrem}_{v_q} \mathbb{E} \left\{ \int_0^T \left( \frac{m}{2} v_q^2(t) -
    V(x(t)) \right)dt + \Phi_T(x(T))\right\}\;. \label{Jq}
\end{equation}
The choice of a complex valued quantum velocity is a compact way of combining
two real-valued variation principles
\begin{eqnarray}
J_R[u^*,v^*] &=& \max_{u} \min_{v} \mathbb{E}\left\{ \int_0^T dt \left (
      \frac{m}{2}(v^2(t)-u^2(t)) -V(x(t))\right ) + S_o(x(T))\right\}\\
J_I[u^*,v^*] &=& - m\, \max_{u} \min_{v} \mathbb{E}\left\{\int_0^T dt\, 
                             v(t) u(t) + R_0(x(T))\right\}\;.
\end{eqnarray}
The first one is a saddle-point action principle, the second one a
saddle-point entropy production principle \cite{Pavon-1995}. The
Euler-Lagrange equation of the quantum Hamilton 
principle is Nelson's stochastic Newton law, the Hamilton-Jacobi-Bellmann
equations of this principle are the Madelung equations. Using stochastic
optimal control theory, we derived the quantum Hamilton equations
\begin{eqnarray}
  dx(t) &=& (v(x(t))+u(x(t)))dt + \sqrt{\frac{\hbar}{m}} dW_f(t)\\
  dp(t) &=& - \frac{d}{dx}V(x) dt + 2\sqrt{\hbar m}\left(\frac{d}{dx} u(x) \right)dW_b(t)\;.
\end{eqnarray}
with $p(t) = m[v(x(t)) + u(x(t))]$ from this
principle \cite{Jeanette1,Michael-Spin} as coupled forward-backward stochastic
differential equations. By now we thus have a structure of
quantum analytical mechanics paralleling that of classical analytical
mechanics. It is the mechanics of time inversion invariant continuous paths in the configuration space
of a system generated by It{\^o} stochastic equations.

All physical observables in this formulation of quantum theory are represented
by stochastic functions on configuration space. Particles possess at each
instant in time position, momentum (a stochastic function of position),
angular momentum, energy and so forth. The expectation values of these agree
with what one calculates based on operator theory. In fact, as far as
accounting for the result statistics goes, quantum analytical mechanics is
equivalent to Hilbert space quantum mechanics, which is just its
Hamilton-Jacobi version. However, representing physical observables as
stochastic variables additionally assigns trajectories of individual particles
a physical reality. This allows for the determination of the duration 
of processes \cite{Jeanette2}, and the modelling of measurement processes
\cite{Michael-Found}. A simulation of the Einstein-Podolski-Rosen-Bohm thought
experiment within quantum analytical mechanics describes the measurement
process in the Stern-Gerlach setups and even reproduces the violation of
Bell's inequality, when performed for an entangled state \cite{Michael-Found}.
An equivalent way of arriving at this structure is to consider the
differential geometry of diffusion processes on manifolds, combining the
second order differential geometry necessitated by the irregularity of the
diffusion paths with the martingale theory of diffusions
\cite{Kuipers-2023,Huang-Zambrini-2023}. 
\begin{table}[h]
\[
  \begin{array}{|c|c|c|}\hline
    ~ & \text{\bf classical analytical mechanics} & \text{\bf quantum analytical mechanics}\\ \hline
    \text{model} & \text{differentiable paths on manifolds} & \text{continuous paths on
                                                 manifolds}\\ \hline
    \text{Newton} & \begin{array}{l}dx(t) = v(t) dt \\ F(x(t)) = m a(t) \end{array}
      & \begin{array}{l} dx(t) = [v(x(t))+u(x(t))]dt + \sqrt{\frac{\hbar}{m}} dW_f(t) \\dx(t) =
    [v(x(t))-u(x(t))]dt + \sqrt{\frac{\hbar}{m}} dW_b(t) \\ F(x(t)) = m
          E[a(t)]\end{array}\\ \hline
    \begin{array}{c}\text{Hamilton} \\ \text{principle}\end{array} & S[x] =
                                \int_{t_0}^{t_1}\left(\frac{1}{2}mv^2(t) -
                                V(x(t))\right)dt & \begin{array}{l} J[v,u] = \mathbb{E} 
                                                   \int_{t_0}^{t_1}\left(\frac{1}{2}m(v(x(t))
                                                   - i u(x(t))^2 \right.\\
                                                     \left. -
                                                     V(x(t))\right)dt \end{array}\\\hline
    \begin{array}{c}\text{Hamilton-} \\ \text{Jacobi} 
     \end{array}& \begin{array}{l} \partial_t\rho(x,t) +
                                    \partial_x\left[\rho(x,t)v(x,t))\right) = 0 \\
                                    v(x,t) = \frac{1}{m} \partial_x S(x,t)\\
                                    \partial_t S(x,t) + 
                                    \frac{1}{2m}\left(\partial_x(S(x,t)\right)1^2
                                    + V(x) =0 \end{array} & \begin{array}{l} \partial_t\rho(x,t) +
                                    \partial_x\left[\rho(x,t)v(x,t))\right) = 0 \\
                                    v(x,t) = \frac{1}{m} \partial_x S(x,t)\\
                                    \partial_t S(x,t) + 
                                    \frac{1}{2m}\left(\partial_x(S(x,t)\right)^2
                                    + V(x) - \frac{\hbar^2}{2m}
                                    \frac{\partial_x^2\sqrt{\rho}}{\sqrt{\rho}}=0 \end{array}
    \\\hline
    \begin{array}{c}\text{Hamilton}\\ \text{equations}\end{array} & \begin{array}{c} dx(t) = \frac{p(t)}{m} dt \\ dp(t) = -
                        \frac{dV(x)}{dx} dt \end{array} & \begin{array}{c} dx(t)
                      = \frac{p(t)}{m} dt + \sqrt{\hbar}{m} dW_f(t)\\
                      dp(t) = \frac{dV(x)}{dx}dt + 2\sqrt{\hbar
                        m}\partial_xu(x,t) dW_b(t)\end{array}\\\hline                                      
  \end{array}
\]
  \caption{Comparison of the different formulations of classical and quantum
    analytical mechanics.}\label{tab1}
\end{table}

Table~\ref{tab1} compares quantum analytical mechanics to classical analytical
mechanics. It is written for simpliticy for one particle in one dimension
but is equally valid for $N$ particles in three dimensions (mutatis mutandis).
For the quantum analytical mechanics, the row titled Hamilton-Jacobi
formulation consists of the Madelung equations, but with the definition
$\psi(x,t) =\sqrt{\rho(x,t)} \exp\{i S(x,t) / \hbar\}$ these equations lead to
the Schrödinger equation \cite{Paul-2026}.

\section{The Dynamics of Measurement}
In this section we want to discuss two examples of a dynamic description of 
measurements based on quantum analytical mechanics. The first one concerns the
description \cite{Kai-2} of a recent levitation experiment \cite{Aspelmeyer-2021}
where the center of mass motion of a silica sphere of 70~nm diameter could be
cooled down close to its zero point motion. This mesoscopic particle is
confined to a harmonic trap furnished by a laser beam, i.e., it is a quantum
harmonic oscillator. This is a quantum particle who's phase space trajectory
for motions along the focus of the laser beam could be observed with 32~ns
temporal resolution \cite{Aspelmeyer-2021}. Quantum harmonic oscillators can be
described by coherent states
\begin{equation}
    \psi (x,t) = (2\pi x_{zpf}^2)^{-\frac{1}{4}}e^{-\frac{(x-x_{cl}(t))^2}{4x_{zpf}^2} + \frac{i}{\hbar} xp_{cl}(t) - \frac{i}{2\hbar}x_{xl}(t)p_{cl}(t)-\frac{i}{2}\omega t}\label{sch}
\end{equation}
with $x_{zpf}^2= \frac{\hbar}{2m\omega}$ and $(x_{cl}(t), p_{cl}(t))$ being
a phase space trajectory of the classical harmonic oscillator. The drift and osmotic velocities therefore are
\begin{eqnarray*}
    v(x,t)&=\frac{p_{cl}(t)}{m}\\
    u(x,t)&=-\omega(x-x_{cl}(t))
\end{eqnarray*}
resulting for the propagation in space in
\begin{equation}
    dx=\biggl(\frac{p_{cl}(t)}{m} -\omega(x-x_{cl}(t)) \biggr)dt +\sqrt{\frac{\hbar}{m}} dW_f(t).\label{coherent_x}
\end{equation}
For $x_{cl}(t)=0, p_{cl}(t)=0$ we obtain the ground state of the quantum
oscillator, for $x_{cl}(t)=2\sqrt{\langle n \rangle} x_{zpf}\sin(\omega t)$,
$p_{cl}(t) =  2\sqrt{\langle n \rangle} p_{zpf}\cos(\omega t)$ we obtain a
coherent state with an energy content of the classical motion of $E_{cl}=
\langle n \rangle \hbar\omega$. 
\begin{figure}[ht]
\begin{center}
\includegraphics[width=0.8\columnwidth]{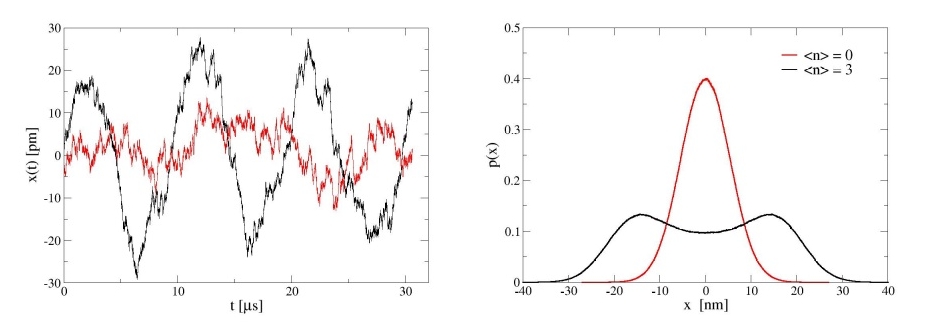}
\end{center}
\caption{Left: sample paths for the ground state and a coherent state with
  energy $3\hbar\omega$ in the classical motion. Right: probability
  distributions for the positions of the particle in these states.}
    \label{coh-paths}
\end{figure}
Two sample paths for choices $\langle n\rangle=0$ (ground state) and $\langle
n\rangle=3$ are shown in Fig.\ref{coh-paths}. The position auto-correlation
function of these paths and its power-spectral density were measured
experimentally \cite{Aspelmeyer-2021} agreeing quantitatively with the predictions
from the coherent state description \cite{Kai-2}. Both trajectories belong to minimum
uncertainty states shown on the right side of Fig.~\ref{coh-paths}. For the
excited coherent state it is a minimum uncertainty path around the classical
trajectory.

This experiment studies a quantum particle which is (almost) visible with a
microscope. It is a particle for all times along its trajectory and it does
not switch between \emph{being a wave spread out in space and then collapsing
  into a particle again}, which one sometimes finds as the interpretation of
wave-particle duality. The wave function or, equivalently, the probability
density and action occurring in the Madelung representation of the wave
function, should never be interpreted as fields in real space. They are fields in
the configuration space of the quantum system. For the one-particle problem
one is tempted to identify configuration space with real space, against which already Schrödinger warned
\cite{Schroedinger-IV}. Probability density and action and with that the wave
function encode our knowledge
about a system, which is determined by the specification of the preparation
procedure. Performing measurements on the system updates this knowledge by
enacting a physical change of the trajectory of the particle.

\begin{figure}[h]
  \begin{center}
    \includegraphics[width=0.75\columnwidth]{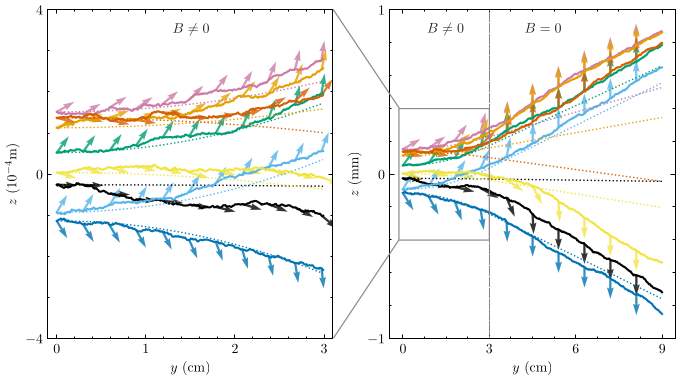}
  \end{center}
\caption{Sample paths of particles bearing an intrinsic magnetic moment in a
  Stern-Gerlach experiment. The first 3 cm, which are enlarged on the left, are
within the magnet, the screen is put at 9 cm distance to the beginning of the
magnet. These dimensions replicate the original experimental setup (taken from
\cite{Michael-Diss} with permission).}
    \label{SG-paths}
\end{figure}
As a second example we want to mention the description of the measurement
process in the Stern-Gerlach apparatus \cite{Michael-Found} using quantum
analytical mechanics. In this case not only the particle position is a hidden
variable but also the particle orientation. The modelling of the experiment
employs the Bopp-Haag picture \cite{Bopp-Haag-1950} of a rotating charged
particle having a magnetic dipole 
moment because of this rotation, as used in the original analysis of the
Stern-Gerlach experiment by Uhlenbeck and Goudsmit \cite{Uhlen-Goud}. The
configuration space of one particle now is given by $\mathbb{R}^3\times SO(3)$,
a Riemannian manifold, and the particle traces a trajectory on this
manifold. Note that the description of particle rotation by a stochastic
process leads to no violation of the speed of light limit, which was Pauli's
original reason to refute such a description. The Quantum Hamilton equations
of this conservative motion process have been mathematically derived and
discussed in the literature \cite{Michael-Spin,Michael-Found}.
Here we only want to make use of one of the results
obtained within this description to elucidate why the use of hidden variables
removes the measurement problem. In Fig.\ref{SG-paths} we show 8 example
trajectories of particles bearing an intrinsic magnetic moment within the
Stern-Gerlach experiment. The particle beam is unpolarized upon entry into the
magnet and has a width of around $150\, \mu$m around the central axes of the
magnet. The full lines with the attached vectors for the orientation of the
particles magnetic moments show the simultaneous deflection and orientation by
the applied inhomogeneous magnetic field. A classical particle would follow
the dotted trajectories creating a stripe pattern on the detector screen. The
quantum trajectories create two spots, one for which the \emph{final}
state of the magnetic moment points up along the z-axis and one for which it
points down. This behaviour is what we are used to encode as spin-up and
spin-down. However, contrary to the standard interpretation, spin here is not
a time-invariant label, but a dynamic degree of freedom (the magnetic moment
of the particle) onto which we act by applying a magnetic field. The
Stern-Gerlach apparatus is designed as an experiment in which a known force
is acting on a known particle property to enact a
physical change of the particles trajectory. Quantum analytical mechanics
offers a way to describe the dynamics of this measurement. The Hamilton-Jacobi
description of this experiment leads to a Schrödinger equation on the
configuration space \cite{Bopp-Haag-1950,Dankel-1970}. In the limit that the
moment of inertia in the rotation energy of the particle goes to zero, this
Schrödinger equation reduces to the spinor description of the Pauli equation
\cite{Bopp-Haag-1950,Wallstrom-1990}, i.e., a description based on the concept of spin as an
internal variable with its associated Hilbert space. The statistics of
measurement outcomes predicted by quantum analytical mechanics  perfectly
agrees with the predictions of the Pauli equation \cite{Michael-Found}, but in
addition one can follow the act of measurement dynamically.

These two applications exemplify why ``hidden'' variables was not a particular
well-suited choice of word. Neither position nor orientation of the particle
are actually hidden in the experiments, they are the only physical entities the
experiments act on. They were ``missing'' from the Hilbert space formulation
of quantum mechanics but are provided by the extension of Hilbert space
quantum mechanics which quantum analytical mechanics provides.

\section{Outlook}
Let me emphasize again, that quantum analytical mechanics is not a replacement
of Hilbert space quantum mechanics but a way to its completion. Hilbert space quantum
mechanics is the Hamilton-Jacobi version of quantum analytical mechanics. All the
myriad results derived from Hilbert space theory over the last 100 years are, of course,
valid. In fact, the crucial test for the other variants that quantum analytical
mechanics could be formulated in, was always that the outcome statistics they
predicted had to agree with the results of the Hilbert space treatment - which
they do. This, in itself, shows that Hilbert space quantum mechanics is not
{\bf the} but {\bf a} complete description of quantum systems. It is the most
highly developed and practical mathematical description of quantum systems,
because it could be reduced to solving a linear partial differential
equation, the Schrödinger equation.

Same as for classical mechanics, the different versions in which quantum
analytical mechanics can be formulated are not completely equivalent. Using
Nelson's original Newtonian version or the quantum Hamilton equations, it
makes perfect sense to calculate and analyze individual trajectories of a
quantum system, not only their average statistics that Hilbert space quantum
mechanics addresses. The most important physical insight this delivers results
from the possibility to analyze the duration of a process like, e.g.,
tunneling \cite{Jeanette2}, whereas in the Hilbert space description there can
not be a Hermition operator for time as an observable. Quantum analytical
mechanics is based on the mathematical encoding of physical observables as
stochastic functions on the configuration space of a quantum system. As such,
position and orientation of a particle become dynamic variables and as these
variables are the ones experiments actually act on, a dynamic description of
the measurement process is possible, removing the measurement problem of
Hilbert space quantum mechanics.

While there have been by now some applications of the non-Hilbert-space formulations
of quantum analytical mechanics to physical problems reported in the
literature, the potential that this broader view of the
mathematics of quantum systems offers has by far not been exploited yet. It is the
goal of this perspective to stimulate further work exploring the possibilities
of this theory for the understanding of quantum systems.

\begin{acknowledgments}
The author acknowledges a long-standing fruitful collaboration with Michael
Beyer, Wilfried Grecksch and Jeanette Köppe. 
\end{acknowledgments}



\bibliography{QM}

\end{document}